\documentclass[aps,prl,showpacs,twocolumn]{revtex4}
\usepackage{graphicx}
\usepackage{amssymb}
\usepackage{amsmath}
\usepackage{bm}
\pdfoutput=1
\begin{document}

\title[]{Observation of Topologically Stable 2D Skyrmions in an Antiferromagnetic Spinor Bose-Einstein Condensate}

\author{Jae-yoon Choi}
\author{Woo~Jin Kwon}
\author{Yong-il Shin}\email{yishin@snu.ac.kr}

\affiliation{Center for Subwavelength Optics and Department of Physics and Astronomy, Seoul National University, Seoul 151-747, Korea}

\date{\today}

\begin{abstract}
   We present the creation and time evolution of two-dimensional Skyrmion excitations in an antiferromagnetic spinor Bose-Einstein condensate. Using a spin rotation method, the Skyrmion spin textures were imprinted on a sodium condensate in a polar phase, where the two-dimensional Skyrmion is topologically protected. The Skyrmion was observed to be stable on a short time scale of a few tens of ms but to have dynamical instability to deform its shape and eventually decay to a uniform spin texture. The deformed spin textures reveal that the decay dynamics involves breaking the polar phase inside the condensate without having topological charge density flow through the boundary of the finite-sized sample. We discuss the possible formation of half-quantum vortices in the deformation process.
\end{abstract}


\maketitle

 Skyrmions are particle-like topological solitons, first proposed to account for the existence of protons and neutrons in high energy physics~\cite{skyrme} and recognized to have an important role in many condensed matter systems such as liquid $^{3}$He-$A$~\cite{He3A}, quantum Hall systems~\cite{QuantumHall}, liquid crystals~\cite{liquidxtal}, and helical ferromagnets~\cite{MnSi,FeCoSi}. Atomic Bose-Einstein condensates (BECs) with internal spin degrees of freedom are attractive for the study of topological objects. The rich structure of their order parameters can accommodate various topological excitations~\cite{HQVRing,Skyrmion,Mizushima,Knots,DiracMonopole,TEreview}. Furthermore, precise spin manipulation techniques have been developed to prepare topological spin structures of interest, providing unique opportunities to study their stability and dynamics~\cite{twocomponentvortex,Kelvinmode,CorelessVortex,DIvortex,SkyrLattice,PolarcoreVortex,2Dskyrmion}.

 Stability is one of the key issues in the study of Skyrmions~\cite{Skyrmion,StableSkyrmion,StableSkyrmion2,SkyrUnstable}. 3D Skyrmions, which are not experimentally discovered yet, are anticipated to be energetically unstable against shrinking to zero size without additional stabilizing mechanisms. In the search for stable Skyrmions, spinor BECs are particularly appealing because of the possibility of engineering experimental conditions such as interaction properties and external potentials including rotation~\cite{StableSkyrmion2}. Earlier experimental efforts have been focused on creating 2D Skyrmion spin textures in spinor BECs by using phase imprinting methods~\cite{CorelessVortex,2Dskyrmion}, where, however, the BECs were not trapped so that the stability of the spin textures could not be investigated. It is important to note that the ferromagnetic phases with $SO(3)$ symmetry studied in the previous experiments~\cite{CorelessVortex,2Dskyrmion} have the trivial second homotopy group, meaning that the spin textures are not topologically protected~\cite{Mueda}. The coreless vortex states with 2D Skyrmion spin textures, referred as Anderson-Toulouse~\cite{ATvortex} or Mermin-Ho vortices~\cite{MHvortex}, have infinite energy in a 2D plane and therefore fundamentally different from the Skyrmions, localized topological solitons with finite energy~\cite{StableSkyrmion2}.

 In this Letter, we demonstrate the creation of 2D Skyrmions in an antiferromagnetic $F=1$ spinor BEC in a polar phase and study their time evolution in a harmonic potential. The order parameter manifold of the polar phase is $M=(U(1)\times S^2)/\mathbb{Z}_2$, having the nontrivial second homotopy group $\pi_2(M)=\mathbb{Z}$~\cite{Mpolar,Mueda}. Thus, the spin textures realized in this work represent the first example of topologically stable 2D Skyrmions in a spinor BEC. Their nature is confirmed by the observation that the spin textures are stable on a short time scale. They are, however, dynamically unstable to deform and eventually decay to a uniform texture. The spatial patterns of the deformed spin textures reveal that the decay dynamics involves breaking the polar phase inside the Skyrmion without having topological charge density flow via the boundary of the condensate. We suggest that half-quantum vortices might be nucleated with the non-polar defects in the deformation process.

 In terms of a unit spin vector $\vec{d}$, the 2D Skyrmion spin texture with $z$-axis symmetry is given as
 \begin{equation}
 \vec{d}(r,\phi)=\cos\beta(r) \hat{z}+\sin\beta(r)\hat{r}
 \end{equation}
 with the boundary conditions, $\beta(0)=0$ and $\beta(\infty)=\pi$. This spin texture has the topological charge
 \begin{equation}
 Q=\frac{1}{4\pi}\int dx dy~\vec{d}\cdot(\partial_{x}\vec{d}\times\partial_{y}\vec{d})=1
 \end{equation}
 which represents the number of times the spin texture encloses the whole spin space.
 The order parameter of a $F=1$ BEC can be written as $\Psi=(\psi_1,\psi_0,\psi_{-1})^\mathrm{T}=\sqrt{n} e^{i\vartheta} \zeta$, where $\psi_m$ is the $|m_z=m\rangle$ component of the order parameter ($m=0,\pm1$), $n$ is the atomic number density, $\vartheta$ is the superfluid phase, and $\zeta$ is a three-component spinor. The ground state of an antiferromagnetic BEC is polar, i.e. $|m_F=0\rangle$~\cite{AFspinor}. Setting $\vec{d}$ as the spin quantization axis, $\zeta=(\frac{-d_x+id_y}{\sqrt2}, d_z, \frac{d_x+id_y}{\sqrt2})^\mathrm{T}$ and the 2D Skyrmion spin texture in a polar BEC is
 \begin{equation}
 \zeta_S(r,\phi) = \left(
 \begin{array}{c} -\frac{1}{\sqrt{2}}e^{-i(\phi)}\sin\beta(r)\\ \cos\beta(r) \\\frac{1}{\sqrt{2}}e^{i(\phi)}\sin\beta(r) \end{array}
 \right).
 \end{equation}
 The spin winding structure around the $z$-axis is expressed with the opposite phase windings of the $|m_z=1\rangle$ and $|m_z=-1\rangle$ components. Note that the total angular momentum of this spin texture is zero.

 \begin{figure}
\includegraphics[width=8cm]{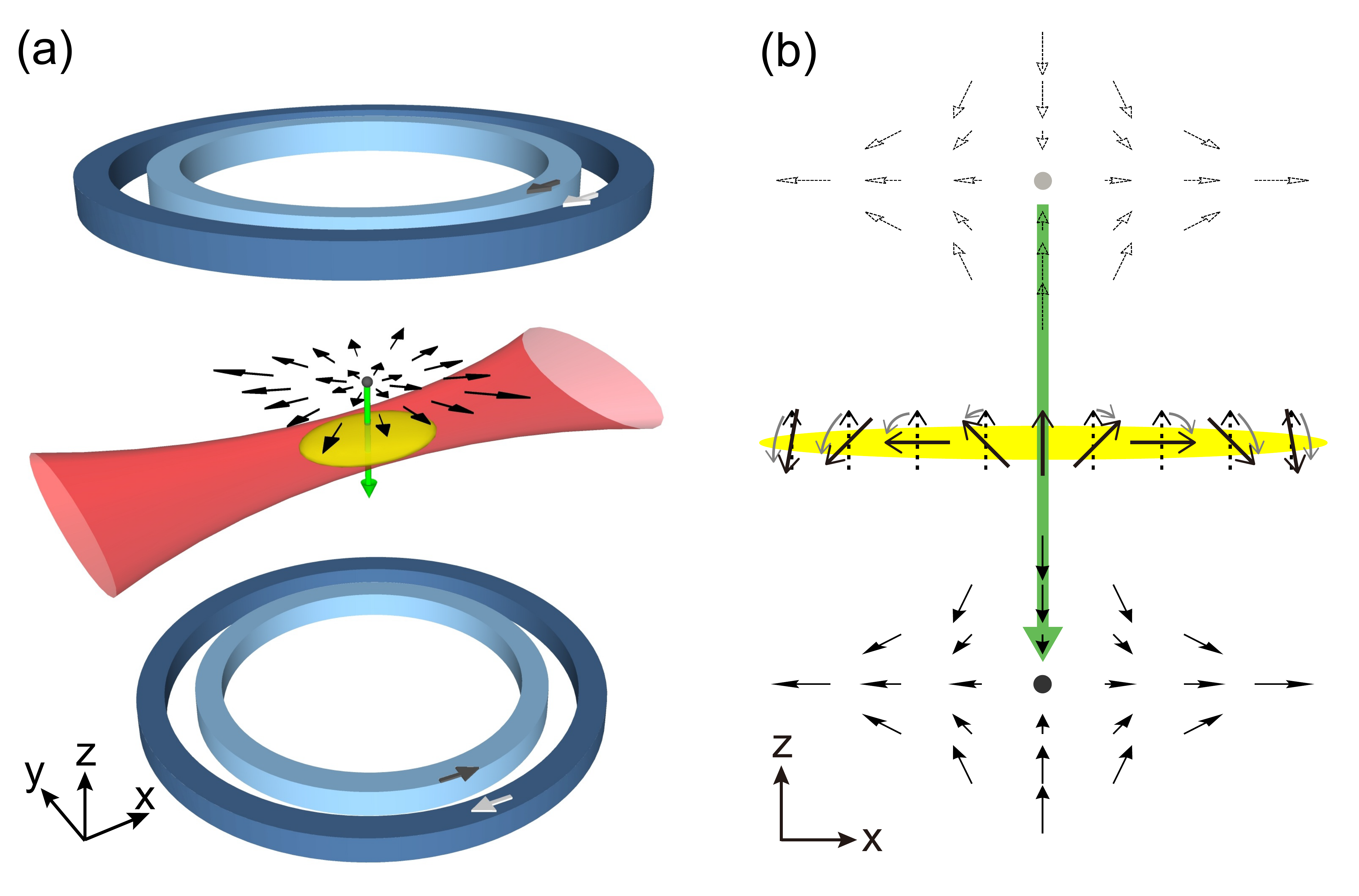}
\caption{
   (color online) Illustration of the creation process of a 2D Skyrmion in a spinor Bose-Einstein condensate. (a) A quasi-2D condensate is prepared in an optical trap and a 3D quadrupole magnetic field is generated from a pair of coils with opposite currents. An additional pair of coils provides a uniform bias field to control the $z$-position of the zero-field center of the quadrupole field. (b) The zero-field point penetrates through the condensate and the atomic spin is driven to rotate by the change of the local magnetic field direction. The tilt angle of the atomic spin depends on its radial position, resulting in a 2D Skyrmion spin texture.}\label{Figure1}
\end{figure}

 For creating a 2D Skyrmion spin texture, we employed a spin rotation method~\cite{CorelessVortex,DIvortex,Isoshima} as illustrated in Fig~\ref{Figure1}. A quasi-2D, polar BEC confined in an optical dipole trap is placed on the $z$=0 plane and a 3D quadrupole magnetic field is generated as
 \begin{equation}
 \vec{B}(r,z) = B'(r\hat{r} - 2z\hat{z}) + B_{z}\hat{z},
 \end{equation}
 where $B'$ is the radial magnetic field gradient. The $z$-position of the zero-field center of the quadrupole field is controlled by the axial bias field $B_z$. Initially, the spin vector of the polar BEC is $\vec{d}=+\hat{z}$ with $B_z\gg B'R > 0$, where $R$ is the radial extent of the conensate. By increasing $B_z$ to $B_z\ll -|B'R|<0$, the zero-field point penetrates through the condensate and the magnetic field on the condensate rotates by $\pi$. In the outer region $R>r\gg r_c \sim [\hbar |\dot{B_z}|/\mu_B B'^2]^{1/2}$, where $\mu_B$ is the Bohr magneton, and $\hbar$ is the Planck constant divided by $2\pi$, the local field direction changes slowly with respect to the local Lamor frequency and $\vec{d}(r)$ adiabatically follows the local field direction to $-\hat{z}$. On the other hand, in the center region $r\ll r_c$, the local field rotates so abruptly that $\vec{d}(r)$ cannot follow the field direction. It is obvious that $\vec{d}(0)$ keeps its direction in $+\hat{z}$. Consequently, the tilt angle $\beta(r)$ of $\vec{d}(r)$ continuously changes from $\beta(0)=0$ to $\beta(R)=\pi$, satisfying the boundary conditions of the Skyrmion spin texture.

\begin{figure}
\includegraphics[width=7.5cm]{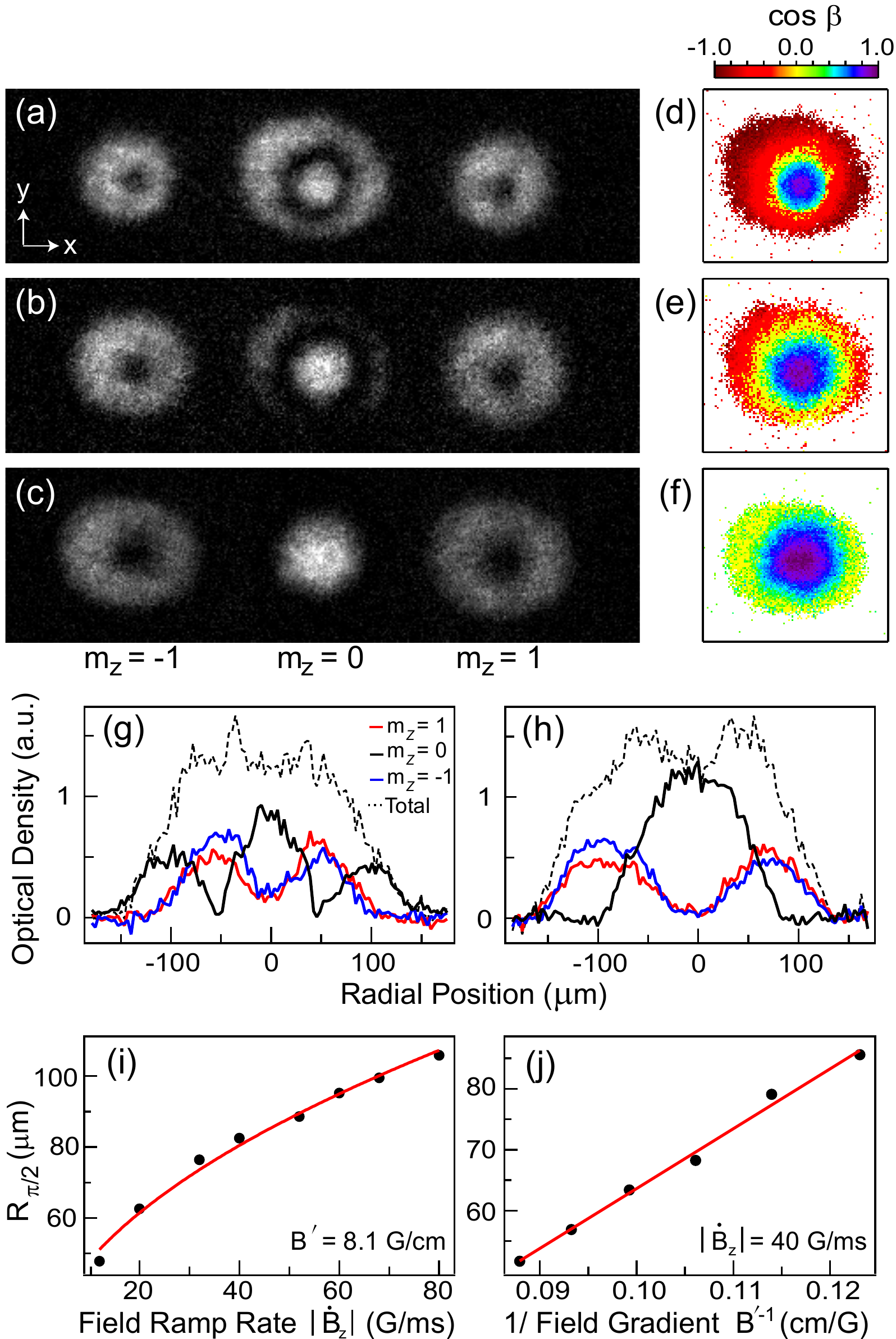}
\caption{
   (color online) Creation of 2D Skyrmions in spin-1 polar Bose-Einstein condensates. (a)$-($c) Density distribution of the $|m_z=0,\pm1\rangle$ spin components were measured by taking absorption images after a Stern-Gerlach separation. The field ramp rate was (a) $|\dot{B}_z|=$12, (b) 32, and (c) 80~G/ms with a quadrupole field gradient $B'=$8.1~G/cm. The contour plots (d)$-$(f) show the distribution of the tilt angle $\beta(x,y)$ in (a)$-$(d), respectively. (g,h) The density profiles of the spin components are the horizontal center cuts in (a,c). The Skyrmion size $R_{\pi/2}$ versus (i) $|\dot{B}_z|$ and (j) $B'^{-1}$. The solid lines are square-root and linear fits to the data points, respectively. The field of view in (a)$-$(c) is 1.2~mm$\times$ 330~$\mu$m.}\label{Figure2}
\end{figure}

 Bose-Einstein condensates of $^{23}$Na atoms were generated in the $|F=1,m_F=-1\rangle$ state  in an optically plugged magnetic quadrupole trap~\cite{SNUBEC} and transferred into an optical dipole trap formed by focusing a 1064-nm laser beam with a 1/$e^2$ beam waist of 1.9~mm(17~$\mu$m) in the $y$($z$)-direction. Further evaporation cooling was applied by lowering the trap depth and a quasi-pure condensate of $1.2\times10^6$ atoms was obtained. The condensate was prepared in the $|m_z=0\rangle$ state by using an adiabatic Landau-Zener RF sweep at a uniform bias field $B_z=21$~G. The trapping frequencies of the final optical trap were ($\omega_x$, $\omega_y$, $\omega_z$)=$2\pi\times$(3.5, 4.6, 430)~Hz and the transverse Thomas-Fermi radii of the trapped condensate were ($R_x$, $R_y$)$\approx$(150, 120)~$\mu$m. For a typical atom density $n=1.2\times10^{13}$~cm$^{-3}$, the spin healing length $\xi_s\approx40~\mu$m~\cite{SpinH} which is much larger than the thickness of the condensate, so the spin dynamics in the condensate is of 2D character.

 For the spin texture imprinting, the quadrupole field was adiabatically turned on to $B'=8.1$~G/cm in 40~ms with $B_z=500$~mG and the the axial bias field was rapidly ramped to $B_z<-500$~mG at a variable ramp rate $\dot{B_z}$. Then the quadrupole field was switched off within 150~$\mu$s and the axial bias field was stabilized to $B_z=-500$~mG within 100~$\mu$s.
 The structure of the spin texture was determined by measuring the density distributions $n_{0,\pm1}(x,y)$ of the $|m_z=0,\pm1\rangle$ components after a Stern-Gerlach spin separation. After switching off the optical trap, the magnetic field was adiabatically rotated and a field gradient was applied for 5~ms in the $x$-direction to spatially separate the three spin components. Then atoms are pumped into the $|F=2\rangle$ state and an absorption image was taken using the $|F=2\rangle\rightarrow |F'=3\rangle$ cycling transition. During the total 15-ms time-of-flight the condensate expanded transversely by less than 10\% and we assume that the measured density distributions adequately reveal the in-situ spin texture.

\begin{figure}
\includegraphics[width=5cm]{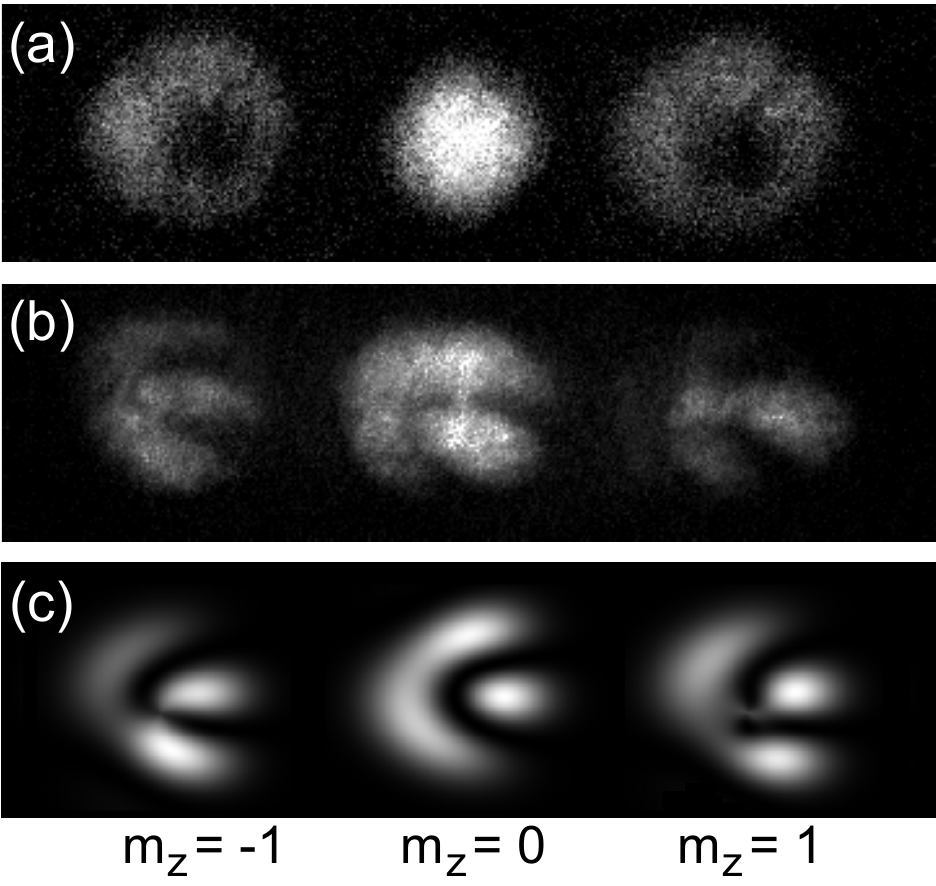}
\caption{
  Interference of the coreless spin vortex state. (a) Initial density distribution of the three spin components. (b) A RF pulse was applied before a Stern-Gerlach separation and fork-shaped interference fringes appear in the $|m_z=0\rangle$ atom cloud. (c) Numerical simulation of our experimental condition with the phase winding numbers $(-1,0,1)$ for the $|m_z=+1\rangle$, $|0\rangle$, and $|-1\rangle$ components.}\label{Figure3}
\end{figure}

 Ring-shaped spin textures were observed in the condensate after the imprinting process (Fig.~\ref{Figure2}). The $|0\rangle$ component shows a clear density-depleted ring and the $|\pm1\rangle$ components occupy the ring region with equal densities. The radius $R_{\pi/2}$ of the ring, characterizing the size of the spin texture, could be controlled with the field ramp rate $\dot{B}_z$ and the quadrupole field gradient $B'$, as $R_{\pi/2}\propto|\dot{B}_z|^{1/2}/B'$, having the same dependence of the critical radius $r_c$ of the adiabatic spin rotation. This validates the aforementioned qualitative description of the imprinting process. The tilt angle $\beta(x,y)$ was reconstructed from $n_{0,\pm1}(x,y)$ with the relation $\cos^2\beta = n_0/(n_1+n_0+n_{-1})$.  When $R_{\pi/2}<0.3 R_x$, $|\cos\beta(R_x)|>0.95$, satisfying the boundary condition of the 2D Skyrmion within our imaging resolution. When $R_{\pi/2}\approx R_x$, the outer ring part of the $|0\rangle$ component disappears and a coreless spin vortex state is formed, where the stationary $|0\rangle$ atoms fill the core of the $|\pm1\rangle$ vortices with opposite circulations~\cite{HalfSkyrmion}

\begin{figure}
\includegraphics[width=5.0cm]{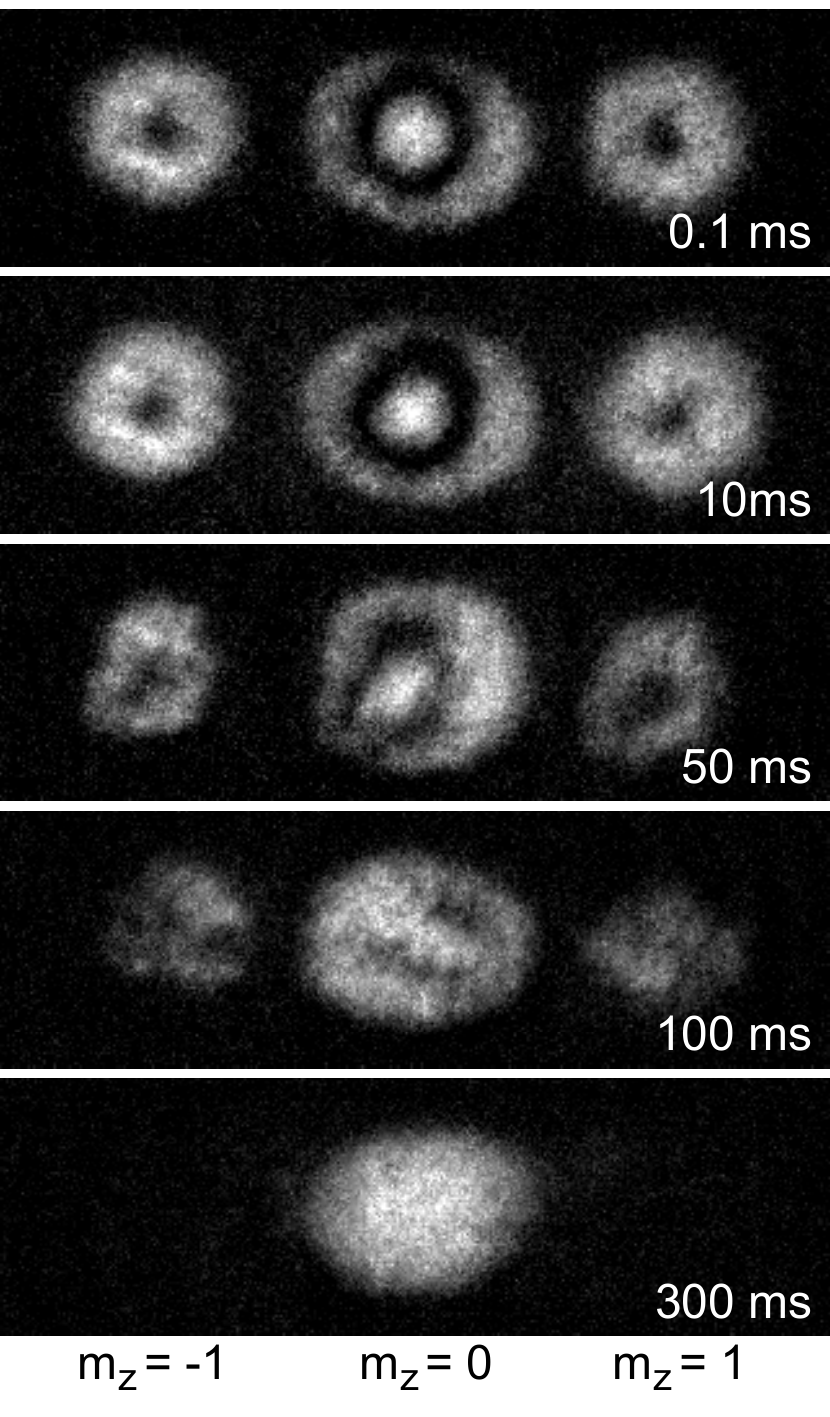}
\caption{
   Deformation and decaying of the 2D Skyrmion in a harmonic potential. The time evolution of the spin texture was measured with a variable dwell time in an optical trap. For this data, $|\dot{B}_z|=$20~G/ms and $B'=$8.1~G/cm. The field of view is 1.0~mm$\times$ 330~$\mu$m.}\label{Figure4}
\end{figure}

 The spin winding structure of the spin texture was confirmed from the matter wave interference between the spin components. A 14-$\mu$s RF pulse of 430~kHz was applied at $B_z$=620~mG before turning off the optical trap. The power and duration of the RF pulse were set with a $|m_z=0\rangle$ condensate to transfer all atoms equally into the $|m_z=\pm1\rangle$ states. Three-to-one, fork-shaped interference fringe patterns were observed (Fig.~\ref{Figure3}), clearly demonstrating that the relative phase winding between the $|\pm1\rangle$ components around the core $|0\rangle$ component is $4\pi$~\cite{vortexinterfere,fork-like}. Since the total angular momentum of the condensate should be conserved to be zero in the imprinting process~\cite{Berry}, the phase winding numbers of the spin components are $(-1,0,1)$. Together with the radial distribution of $\beta$, this clearly shows that a 2D Skyrmion is indeed created in the polar condensate.

 In order to study the stability of the 2D Skyrmion, we measured the time evolution of the spin texture with a variable dwell time $t$ in the optical trap and observed that it dynamically deforms and decays to a uniform spin texture (Fig.~\ref{Figure4}). Modulation of the ring structure gradually developed and distorted the whole spin texture over 100~ms. Eventually, the spin texture became uniform in the $|m_z=0\rangle$ state at $t>300$~ms. Recent theoretical calculations have shown that 2D Skyrmion excitations are energetically unstable to decay to a uniform spin texture by expanding or shrinking in a harmonic potential~\cite{SkyrUnstable} but we estimate that the thermal dissipation is negligible in our experiments for the upper bound of the condensate temperature $<65$~nK. The decay of the Skyrmion is a dynamically driven process.

 It is remarkable that the outer $|0\rangle$ component kept surrounding the $|\pm 1\rangle$ components in the decay dynamics, implying that there was no topological charge density flow via the boundary of the finite-sized condensate. Since it is topologically impossible for the isolated Skyrmion to unwind to a uniform spin texture when the global polar phase is preserved, non-polar defects must have developed inside the condensates during the decay dynamics. Breaking the polar phase might be attributed to the quadratic Zeeman effects and the induced spin currents~\cite{Knots}. The spin relaxation time of a $|m_z=\pm 1\rangle$ equal mixture was measured to be over 1~s at $B_z=500$~mG, clearly indicating that the rapid development of non-polar defects is due to the presence of the Skyrmion spin texture.

\begin{figure}
\includegraphics[width=8.0cm]{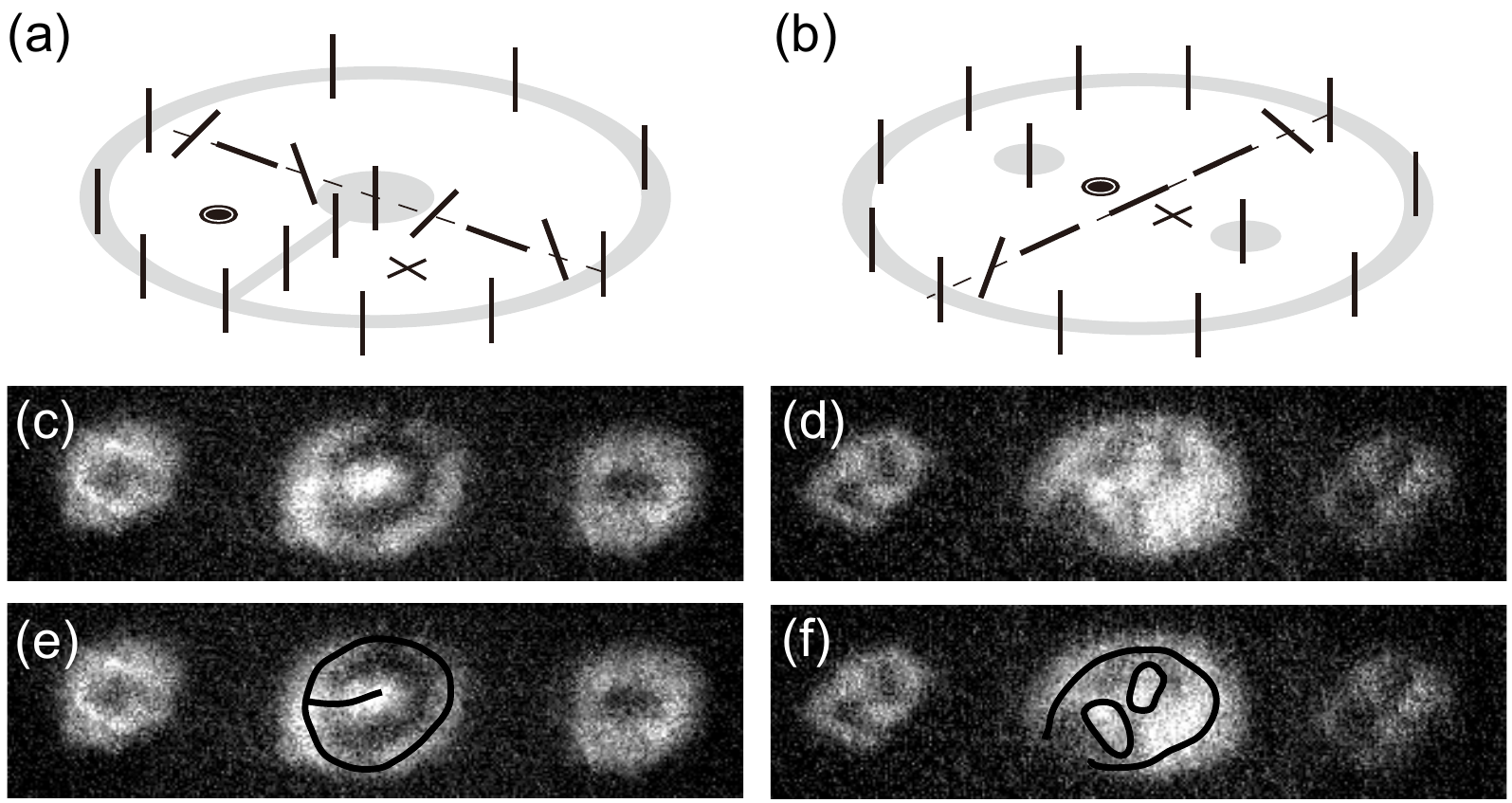}
\caption{Dynamic formation of a pair of half-quantum vortices. The line segments represent the axis of the spin vector $\vec{d}$ and the grey area indicates the region where the $|m_z=0\rangle$ component mainly occupies. The non-polar cores of the half-quantum vortices are marked by $\odot$ and $\times$, implying opposite windings of the superfluid phase. Examples of the dynamic evolution of the 2D Skyrmion spin texture at (c) $t=30$ and (d) 50~ms, displaying density profiles similar to those in (a) and (b). (e,f) Same as (c,d) with guidelines for the peak density regions in the $|m_z=0\rangle$ atom cloud.}\label{Figure5}
\end{figure}

 One interesting observation in the deformed spin textures is that the core $|0\rangle$ component could be connected to the outer component or divided into two parts (Fig.~\ref{Figure5}). Because of the symmetry of the order parameter of the polar BEC under $(\vec{d},\vartheta)\rightarrow(-\vec{d},\vartheta+\pi)$, the two regions with a $\pi$ disclination in $\vec{d}$ can be continuously connected via a $\pi$ change of the superfluid phase. This observation suggests possible formation of half-quantum vortices around non-polar local defects in the deformation process as described in Fig.~\ref{Figure5}. The $|\pm 1\rangle$ components seemed to keep occupying the density-depleted region of the $|0\rangle$ component without developing noticeable local spin polarization $n_1-n_{-1}$, supporting the possibility. A similar situation has been studied with a point defect in a polar 3D BEC, where the 't Hooft-Polyakov monopole continuously deforms to a half-quantum vortex ring (Alice ring)~\cite{Alicering}.

 In conclusion, we have demonstrated the creation of the topologically stable 2D Skyrmions in a polar BEC and presented the first measurement of the dynamics of the Skyrmion in a harmonic potential. This work opens a new perspective for the study of Skyrmion dynamics in spinor BECs. The experimental technique can be extended to create multiple Skyrmions in a condensate or other exotic spin textures such as knots~\cite{Knots} and monopoles~\cite{Alicering} in 3D condensates.

 The authors thank W. Ketterle and J.~H. Han for discussions. This work was supported by National Research Foundation of Korea (NRF) grants (No. 2011-0004539, 2011-0017527, 2011-0001053, and WCU-R32-10045) and Research Settlement Fund for the new faculty of SNU. JC and WJK acknowledge support from Global PhD Fellowship and BK 21 Fellowship, respectively.


\begin{references}
\bibitem{SpinH} $\xi_s=2\pi\hbar/\sqrt{2m c_2 n}$
 , where $m$ is the atomic mass and $c_2=4\pi\hbar^2 (a_2-a_0)/3m$. $a_{0,2}$ are s-wave scattering length for total spin 0, 2 and $a_2-a_0\simeq5.7a_B$ ($a_B$ is the Bohr radius)~\cite{sclength}.
\bibitem{HalfSkyrmion} The vortex state is often referred to as a `half' Skyrmion because $\beta(R_{TF})=\pi/2$, which is not a topologically stable object.
\end{references}
\end{document}